\documentclass[12pt,a4paper]{article}
\usepackage[english]{babel}
\usepackage[utf8]{inputenc}
\usepackage[dvips]{graphics}
\usepackage{longtable}
\usepackage{lscape}
\usepackage{multirow}
\catcode`\{=1
\catcode`\}=2
\def\lref#1{{\tt [#1]}}

\title{GuStL -- An Experimental Guarded States Language}
\author{Oskar Schirmer}
\date{Göttingen, 2016-12-20 \\
~ \\
{\small \it revised 2018-07-10 \\
(reference to BCPL added, minor clarifications)\\}
~ \\
{\small \it revised 2023-06-06 \\
(transmission of strings added)\\}
~ \\
{\small \it revised 2024-09-12 \\
(parameters and transmission of array subranges added)}}

\begin{document}

\maketitle
\vfill

\section*{Abstract}

Programming a parallel computing system that consists of
several thousands or even up to a million message passing
processing units may ask for a language that supports waiting
for and sending messages over hardware channels.

As programs are looked upon as state machines, the language
provides syntax to implement a main event driven loop.

The language presented herewith surely will not serve as
a generic programming language for any arbitrary task.
Its main purpose is to allow for a prototypical implementation
of a dynamic software system as a proof of concept.

\vfill
\newpage

\tableofcontents
\newpage

\section{Introduction}

The {\it Guarded States Language} essentially does combine
well known concepts of existing languages, while
any concept not needed for its main purpose is avoided.
Its basic concept is imperative sequential operations.
Syntactically it is inspired by Algol (\lref{1963nb})
and Pascal (\lref{1975jw}).
Main program execution is organized as a state machine
as with SDL (\lref{2007ir}).
States are subject to guards, similar to those introduced
in \lref{1975ed}.
For channel operations, input and output, the notation
follows \lref{1978ch}.

A software system may consist of various processes
running in parallel,
each of which is a program executed sequentially,
one instruction at a time.
Channels may be
set up dynamically to allow transmission of data,
i.e. messages, from one process to the other.

Depending on the current process state, different guards
may be evaluated, causing some statement sequence
to be executed as soon as the condition of the corresponding guard is met.

Statements serve to implement traditional imperative
operations, such as assigments, loops, conditional
execution and subroutine invocation.

The language does only provide features restricted
to the processing unit the program is executed on.
As a consequence, and as dynamic process creation
does imply activity of various system processes
-- especially in the case of creating a remote process --,
the language does not provide syntactic means
to describe instantiation of a new process.

\section{Syntax}

A program is made up of a sequence of symbols.
A symbol may be one of: special symbol, reserved word,
identifier, number, or string.

Symbols may be separated by white space,
i.e. blank, tabulator, or newline characters.
Any character sequence starting with a left brace
up to and including the next right brace
is considered white space, too,
as long as it does not start within a string.
The latter allows insertion of comments.
Note, that nested comments are not valid.

For the special symbols and reserved words,
see the formal syntax definition below.
Reserved words are to be written in lower case.

An identifier is composed of a sequence of upper and
lower case letters, digits, and the underline character.
It must not start with a digit. Identifiers are case
sensitive.

Numbers are given in either decimal, hexadecimal,
or character notation.
For decimal notation a number is written as a sequence
of digits, for hexadecimal notation, it is preceded by {\tt 0x}.
A number given in character notation is the characters
implicit textual representation\footnote{
It is assumed that the usual implicit textual representation for characters
is its Unicode code point as defined in \lref{2009uc}}
enclosed by single quotes.
Scientific notation is not supported.

A string is a series of characters
-- i.e. their implicit textual representations --
enclosed by double quotes.
To include a double quote in a string, it is duplicated.
A string is encoded as an array at one word per character.

A valid sentence for a program must match the following
partially formalised syntax description.\footnote{
For reasons of simplicity, it is given in a somewhat modified
version of the notation inspired by P$\bar{a}\d{n}$ini (\lref{1840ob})
and initially used by Backus (\lref{1959jb}).
To achieve a fully
formalised definition, including semantics,
one would need to describe the language by using, e.g.,
an extended affix grammar (\lref{1997dw}).
The scope of identifiers is hinted with a subscript $_{!scope}$,
where identifiers are declared, and with a subscript $_{?scope}$,
where they are assumed to be already declared in the scope given}

\vspace*{2ex}

\noindent
$process ::=$
  \{ $constdef$ $\vert$ $wordsdecl$ $\vert$ $proceduredecl$ $\vert$ $functiondecl$ \}\\
\hspace*{8ex}
  ``{\tt process}'' $ident_{!process}$ ``{\tt (}'' $ident_{!port}$ $\lbrack$ ``{\tt ,}'' $ident_{!const}$ $\rbrack$ ``{\tt )}''\\
\hspace*{8ex}
  $\lbrack$ $statesdecl$ $\rbrack$
  $\lbrack$ $portsdecl$ $\rbrack$\\
\hspace*{8ex}
  \{ $constdef$ $\vert$ $wordsdecl$ $\vert$ $proceduredecl$ $\vert$ $functiondecl$ \}\\
\hspace*{8ex}
  ``{\tt start}''
  \{ $statement$ \}
  \{ $guardedstate$ \}
  ``{\tt stop}''\\
$statesdecl ::=$
   ``{\tt state}'' $ident_{!state}$ \{  ``{\tt ,}'' $ident_{!state}$ \}\\
$portsdecl ::=$
   ``{\tt port}'' $ident_{!port}$ \{  ``{\tt ,}'' $ident_{!port}$ \}\\
$constdef ::=$
  ``{\tt const}'' ( $ident_{!const}$ ``{\tt =}'' $expression$ $\vert$
  $ident_{!constarray}$\\
\hspace*{8ex}
  ``{\tt \lbrack}'' $\lbrack$ $expression$ $\rbrack$ ``{\tt \rbrack}'' ``{\tt =}'' $constarray$ \{  ``{\tt ,}'' $constarray$ \} )\\
$constarray ::=$
  $expression$ $\vert$ ``{\tt "}'' $chars$ ``{\tt "}''\\
$wordsdecl ::=$
  ``{\tt word}'' $worddecl$ \{ ``{\tt ,}'' $worddecl$ \}\\
$worddecl ::=$
  $ident_{!word}$ $\vert$
  $ident_{!array}$ ``{\tt \lbrack}'' $expression$ ``{\tt \rbrack}''\\
$proceduredecl ::=$
  ``{\tt procedure}'' $ident_{!procedure}$ ``{\tt (}'' $formalparameters$ ``{\tt )}''\\
\hspace*{8ex}
  \{ $wordsdecl$ \}
  ``{\tt do}''
  \{ $statement$ \}
  ``{\tt return}''\\
$functiondecl ::=$
  ``{\tt function}'' $ident_{!function}$ ``{\tt (}'' $formalparameters$ ``{\tt )}''\\
\hspace*{8ex}
  \{ $wordsdecl$ \}
  ``{\tt do}''
  \{ $statement$ \}
  ``{\tt return}'' $expression$\\
$formalparameters ::=$
  $\lbrack$ $formalparameter$ \{ ``{\tt ,}'' $formalparameter$ \} $\rbrack$\\
$formalparameter ::=$
  ``{\tt const}'' $ident_{!constarray}$ ``{\tt \lbrack}'' $\lbrack$ $expression$ $\rbrack$ ``{\tt \rbrack}'' $\vert$\\
\hspace*{8ex}
  $ident_{!array}$ ``{\tt \lbrack}'' $\lbrack$ $expression$ $\rbrack$ ``{\tt \rbrack}'' $\vert$
  $ident_{!word}$ $\vert$
  ``{\tt port}'' $ident_{!port}$\\
$guardedstate ::=$
  ``{\tt on}'' $statelist$ $\lbrack$ ``{\tt \textbackslash}'' $guard$ $\rbrack$ ``{\tt :}''
  \{ $statement$ \}\\
$statelist ::=$
  $ident_{?state}$ \{ ``{\tt ,}'' $ident_{?state}$ \}\\
$guard ::=$
  $ident_{?port}$ ``{\tt !}'' $\vert$ $reception$ $\vert$ $expiration$\\
$statement ::=$
  $conditional$ $\vert$
  $repetition$ $\vert$
  $transition$ $\vert$
  $assignment$ $\vert$\\
\hspace*{8ex}
  $transmission$ $\vert$
  $procedurecall$ $\vert$
  $inlineasm$\\

\noindent
$conditional ::=$
  ``{\tt if}'' $expression$ ``{\tt then}'' \{ $statement$ \}\\
\hspace*{8ex}
  \{ ``{\tt elseif}'' $expression$ ``{\tt then}'' \{ $statement$ \} \}\\
\hspace*{8ex}
  $\lbrack$ ``{\tt else}'' \{ $statement$ \} $\rbrack$
  ``{\tt done}''\\
$repetition ::=$
  $\lbrack$ ``{\tt while}'' $expression$ $\rbrack$
  ``{\tt repeat}'' $expression$ ``{\tt times}''\\
\hspace*{8ex}
  \{ $statement$ \}
  ( ``{\tt done}'' $\vert$ ``{\tt until}'' $expression$ )\\
$transition ::=$
  ``{\tt next}'' $ident_{?state}$\\
$assignment ::=$
  $variable$ ``{\tt :=}'' ( $expression$ $\vert$ $inlineasm$ )\\
$transmission ::=$
  $ident_{?port}$ ``{\tt !}'' $transdata$\\
$transdata ::=$
  ( $parameter$ $\vert$ ``{\tt "}'' $chars$ ``{\tt "}'' ) $\lbrack$ ``{\tt ,}'' $transdata$ $\rbrack$\\
\hspace*{8ex}
  $\vert$ ``{\tt end}'' $\vert$ ``{\tt pause}''\\
$reception ::=$
  $ident_{?port}$ ``{\tt ?}'' ( $variable$ $\vert$ ``{\tt end}'' )\\
$variable ::=$
  $ident_{?word}$ $\vert$ $ident_{?array}$ ``{\tt \lbrack}'' $expression$ ``{\tt \rbrack}'' $\vert$ $ident_{?port}$\\
$expression ::=$
  $simpleexpr$\\
\hspace*{8ex}
  $\lbrack$ ( ``{\tt =}'' $\vert$ ``{\tt <>}'' $\vert$ ( ``{\tt <}'' $\vert$ ``{\tt >}'' $\vert$ ``{\tt <=}'' $\vert$ ``{\tt >=}'' ) $\lbrack$ ``{\tt \$}'' $\rbrack$ ) $simpleexpr$ $\rbrack$\\
$simpleexpr ::=$
  $term$ \{ ( ``{\tt +}'' $\vert$ ``{\tt -}'' $\vert$ ``{\tt |}'' $\vert$ ``{\tt \^{}}'' $\vert$ ``{\tt or}'' ) $term$ \}\\
$term ::=$
  $factor$\\
\hspace*{8ex}
  \{ ( ``{\tt *}'' $\vert$ ``{\tt \&}'' $\vert$ ``{\tt and}'' $\vert$ ``{\tt <<}'' $\vert$ ( ``{\tt /}'' $\vert$ ``{\tt \%}'' $\vert$ ``{\tt >>}'' ) $\lbrack$ ``{\tt \$}''  $\rbrack$ ) $factor$ \}\\
$factor ::=$
  $\lbrack$ $unaryop$ $\rbrack$
  ( $variable$ $\vert$ $constant$ $\vert$ $functioncall$ $\vert$ ``{\tt now}''\\
\hspace*{8ex}
  $\vert$ $reception$
  $\vert$ ``{\tt (}'' $expression$ ``{\tt )}'' )\\
$unaryop ::=$
  ``{\tt -}'' $\vert$ ``{\tt \textasciitilde}'' $\vert$ ``{\tt not}''\\
$constant ::=$
  $number$ $\vert$ ``{\tt '}'' $char$ ``{\tt '}'' $\vert$
  ``{\tt \#}'' ( $ident_{?array}$ $\vert$ $ident_{?constarray}$ ) $\vert$\\
\hspace*{8ex}
  $ident_{?const}$ $\vert$
  $ident_{?constarray}$ ``{\tt \lbrack}'' $expression$ ``{\tt \rbrack}''\\
$procedurecall ::=$
  $ident_{?procedure}$ ``{\tt (}'' $actualparameters$ ``{\tt )}''\\
$functioncall ::=$
  $ident_{?function}$ ``{\tt (}'' $actualparameters$ ``{\tt )}''\\
$actualparameters ::=$
  $\lbrack$ $parameter$ \{ ``{\tt ,}'' $parameter$ \} $\rbrack$\\
$parameter ::=$
  $expression$ $\vert$ ( $ident_{?array}$ $\vert$ $ident_{?constarray}$ )\\
\hspace*{8ex}
  $\lbrack$ ``{\tt \lbrack}'' $expression$ ``{\tt :}'' $\lbrack$ $expression$ $\rbrack$ ``{\tt \rbrack}'' $\rbrack$\\
$expiration ::=$
  ``{\tt after}'' $expression$\\
$inlineasm ::=$
  ``{\tt asm}'' $singleasm$ \{ ``{\tt ,}'' $singleasm$ \}\\
$singleasm ::=$
  $\lbrack$ $unaryop$ $\rbrack$ $constant$ $\vert$ ``{\tt (}'' $expression$ ``{\tt )}''

\section{Semantics}

The process is a state machine, based upon an arbitrary
number of states. As a special case, the number of states
may be zero.
In any other case, all state identifiers must be declared
next to the process header.

To allow external communication, a process may want to
transmit or receive data over channels.
To set up a channel at runtime,
it must be dynamically assigned to a port,
which in turn is static.
Any port is referred to by an identifier
declared before being used.

Constant identifiers may be declared to stand for
constant words or constant arrays.
The value of a constant array is given as a list
of numbers, where all or parts of the list may be
replaced by a string.

The basic unit of data to operate on is a word.
It represents a natural or integer number, which depends on
which operation is performed on it.
Different basic types to restrict specific operations
or to determine details of an operation -- like signedness --
do not exist.\footnote{Various programming languages
define signedness as a property of each single data unit.
This way, wherever data units of different type are subject
to an operation, implicit conversion rules need to be applied,
an everlasting source of confusion.
To avoid this mess, GuStL makes signedness
be a property of the operation itself.
Much the same approach had been chosen earlier in the
design of the programming language BCPL, see \lref{1980rw}}
A variables identifier may
either be declared to refer to a single word, or,
when declared as an array, to refer to a sequence of words.
Variables must be declared prior to being used.

Subroutines -- both functions and procedures --
may be declared to implement sequences of statements
for multiple use in the program.
Upon termination, functions provide a word return value
to be used in the invocating expression,
while procedures do not return any value.
Subroutines may be declared to take a list of parameters,
which may be words or arrays or ports.
Furthermore,
subroutines allow declaring local single word variables,
which are instantiated at subroutine invocation
and cease at the end of the subroutines execution.
Subroutines must be declared prior to being used.

Statements in the program or in a subroutine are
executed sequentially, one by one.
For a conditional statement, a series of expressions
is evaluated, until one is met -- i.e. the expression
evaluates to non-zero -- and the corresponding
list of statements is executed.
If no condition is met and an alternative branch is
given with {\tt else}, that alternative branches
statements are executed.
A repetition statement is executed for a maximum number
of iterations.
Furthermore, it is not iterated or not executed at all,
when the initial condition is not met,
and it is no longer iterated
when the terminating condition is met.

The right side of an assignment is evaluated and
the result is assigned to the word or port
on the left side. A value assigned to a port takes
it for a destination port number and prepares a channel
to be set up from the local to the destination port.

For channel data transmission, the
port to communicate over is given on the left side
of the statement.
The data to transmit is given on the right side,
a list of arbitrary values,
optionally followed or replaced by
a special token to end or pause a message.
Parts or all of the list of values may be
replaced by array subranges.

Channel data reception is not a statement,
but a factor in an expression.
As with transmission, the port to receive data from is
given on the left side, while on the right side is either
a variable to store data into or the end symbol.
Both variants consume an incoming end token when it is
available, but if data is pending, it is only consumed
in the first case and stored into the variable.
Same as with the assignment, storing the newly received word
to a port sets the ports destination.
The factor evaluates to non-zero whenever the expected
token is received, which is data for the first variant
and an end token for the second.
Otherwise, it evaluates to zero.

Calling a subroutine implies evaluation of the list
of actual parameters, which must match the list of
formal parameters given in the subroutines declaration.
For formal word parameters, the corresponding
expression is evaluated and its result used in the
subroutine (call by value).
For formal array parameters, the reference to the corresponding
variable or array subrange is determined and used in the subroutine
(call by reference).
For formal port parameters, the given port identification
is used in the subroutine (call by reference).
For a subroutine with more than one parameter
the order of parameter evaluation is not defined.

An array subrange may be an array in its entirety
given by its identifier.
To refer to a single element of an array, the identifier
is followed by an index in brackets.
Where the index is followed by a colon,
the subrange from that position up to the end of the array is denoted.
Finally, a length may by given next to the colon
to restrict the size of the subrange.

The transition statement is used to change the machines state.
It must not be used in a subroutine.
Additionally, every execution path in the main program
must end in a transition statement.

The main program starts with a list of statements for
initialisation purposes, and ends with a list of
statements to handle proper process termination.
In between, an arbitrary number of guarded statements
is provided. Whenever the initialising list of statements
or another guarded statement of the program
has finished executing the next guarded statement may
be selected for execution. Only those guarded statements,
that match the current state, are considered.
From these, one random guarded statement is selected,
for which the condition of its guard is met.
When no condition of any guard is met at all,
the process is stalled until at least one condition is met.

A guard does either check a port for transmission readiness,
for availability of data to receive, for an end token
received on a port, or for a given time being expired.
When data is available and the corresponding
guarded statement is selected for execution, the data
received is actually stored into the variable given.
If for a state a guard is declared to check for availability
of input data, another guard to check for an end token
on the same port is obligatory for the same state.
For one state several different guards may be declared,
provided they do not check for the same condition on
the same port, and that no two guards check for expiration.

By default, when evaluating an expression, all operations
are performed as unsigned. For various operators, however,
appending a {\tt \$} sign to the operator itself
may be used to ask for signed operation.
Signed division is defined to be Euclidean (see e.g. \lref{2001dl}).
Any expression is composed of one or two simple expressions,
simple expressions are composed of terms,
terms are composed of factors.
A factor may be a single word or a word from an array,
a constant number or an element from a constant array,
the size of an array,
the call of a function -- using its return value,
the current time ({\tt now}),
the result of checking an input port for having received an end token,
or another expression.
Any factor may be prepended by a unary operator.
The order in which parts of an expression are evaluated is not defined.

To allow machine specific operations,
inline assembly is allowed both as a statement
and as the right side of an assignment.
Inline assembly is given as a list of opcodes,
but may include also ordinary expressions to be evaluated.
Opcodes depend on the implementation, see section \ref{compiler}.
An expression included in inline assembly will leave
the corresponding result on the data stack.

\section{Process Creation}

For traditional single node computers according to the
von-Neumann architecture (\lref{1945jn}),
the basic resource to be allocated by processes at runtime is memory,
taken from a uniform pool of contiguous memory words.

For machines made up of large amounts of parallel processing units,
each with its own local memory,
the allocatable quantity is a single processing unit.
Thus, allocating resource does no longer mean
to reserve a portion of memory, but to start a process
on a previously idle processing unit: {\it process creation}.

The program to run on the allocated
processing unit has to be identified, e.g. by name,
which in turn may be variable, and does not refer to
data held inside the current program, so for the language
to natively support process creation a hypothetical rule
could look like the following -- where the identifier
would refer to a string denoting the new processes name:\\

\noindent
$creation ::=$
  ``{\tt new}'' $ident_{?array}$ ``{\tt (}'' $expression$ ``{\tt )}''\\

However, depending on the hardware, starting a process may
highly depend on supportive software, i.e. the operating system.
The compiler would need to produce operating system dependent
code for process creation.
To avoid complexity and increase flexibility,
means for process creation shall instead be provided
by the operating system as an ordinary subroutine.
An outline on how it could be declared is given hereafter.

Once the new process is being executed, the process that started
it may need to communicate with it. To do so, it needs to know
at least one port of the new process to attach to, the
{\it control port}. The subroutine for process creation will
return its global identification, so it must be a function.
For the new process, the port is declared as the first
parameter to the process declaration header.

As the processing unit and its local memory are -- once allocated --
an indivisible unit, the amount of memory available during
lifetime of the process cannot be changed. Only at the time
of creation of the process, the amount of memory given to it
may be chosen. For this purpose, the subroutine to create the
new process not only needs to know its name, but also a number,
which will be available inside the new process as a constant,
the {\it dimension}. \label{dimension}
To make it usable in a running program,
it is declared to be a constant as the second parameter
to the process header. It is optional and defaults to zero.

Hardware driver processes may need to access specific
hardware locations, so it is crucial to start such a process
on a given processing unit.
For other reasons, it may be desirable to choose
the processing unit from a restricted subset of all units
available, so some flags might possibly be helpful.
An additional parameter to the
function starting the process is used to commit the information
needed. The function may be declared as follows:

\begin{verbatim}
function new(name[], dimension, extra)
\end{verbatim}

\section{Segmented Program Development}

When writing a new program, parts of it may be similar or
even identical to portions of a previously implemented program.
The wish to reuse these portions and avoid repeated work
should represent one of the basic desires of every programmer.

The most simple solution would be to simply {\it copy and paste}
the portion in question to the new program. While in earlier
days this was frowned upon for the reason of wasting storage,
today the untraceable propagation of programming mistakes
is a major concern.

Separate compilation units had the advantage to circumvent
the need to recompile all of the code used in a program
by keeping the program divided into portions even in
its compiled state.
This was useful in times of slow compiling machines,
but bears the risk of binary incompatibilities when
interfaces are changed in source files.
Further, an additional {\it linker} program or a linking loader
is needed to enable program execution.

Assembly of the program at compile time may be achieved
using {\it include files}, a method
to splice a set of source files.\footnote{Using
C (\lref{1978kr}) usually involves a combination
of both linkable program portions, called {\it libraries},
and include files for interface declaration}
But it may also lead to confusion of which include file
is used where, especially when using nested inclusions.

To avoid the latter problem, source file composition
shall be restricted to simple concatenation.
In fact, neither the language nor the compiler need
to take any provisions for it, as the concatenation
may well be performed by the invoking entity before
feeding the complete source program into the
compiler itself.

The fact that insertion of one source file into another
is impossible makes it necessary to allow
declarations before the process declaration of
a program, as generic declarations need to be prepended
to the main program.

\section{Compiler}
\label{compiler}

A first compiler for GuStL has been implemented to
produce code for the experimental machine {\it NOP}
(\lref{2016os}).
The compiler itself is implemented in C,
running in a Unix environment.
Its single pass design follows the basic concept of \lref{1977nw}.
It compiles source from {\it standard input}
to binary code on {\it standard output}.
Matching the target machine, words are implemented
as 32 bit wide,
least significant byte first,
and signed numbers are represented in two's complement (\lref{1982bg}).
The implicit textual representation for characters
is its Unicode code point as defined in \lref{2009uc},
encoded using UTF-8 in the source file (see also \lref{2003fy}).

The binary code itself is preceded by a six word header
to allow direct use in an operating system environment:

~

\begin{tabular}{|p{.6in}|p{4.3in}|} \hline
magic & {\tt 0x85cf80cf} \\ \hline
flags & {\tt 0} \\ \hline
$d_1$ & \multirow{2}{3.9in}{data memory size:
ld$_1$-ld$_0$ = $d_1 \cdot dimension + d_0$,
for {\it dimension} see page \pageref{dimension},
see also \lref{2016os}} \\ \cline{1-1}
$d_0$ & \\ \hline
code & code and constants memory size: lc$_1$-lc$_0$,
see \lref{2016os} \\ \hline
entry & word offset to initial instruction, relative to lc$_0$,
see \lref{2016os} \\ \hline
\end{tabular}

\section{Discussion}

The main purpose of the programming language GuStL
is to allow for a prototypical implementation
of a dynamic software system,
and its design is restricted to features that are needed
to serve this purpose.

There is only very limited support for string handling,
namely definition of constant arrays.
It might be useful to provide more support for string handling,
e.g. transmission and reception of sequences instead of a single
word.
However, the latter might reduce program flow determinism,
because time behavior for the processing of a sequence
potentially is more complex than for a single word,
scattering non-determinism from controlling guards
into imperative program flow.

Direct recursion in a subroutine is supported,
but recursion involving several subroutines is not,
as there is neither nested nor forward declaration of subroutines.
With local memory being quite limited,
it is arguable whether system abstraction
is adequate to justify support of massive recursion.

Likewise, there is no specific support for
floating point calculations, list processing,
bit field computation or set arithmetics (as defined for Pascal),
data pointers or complex data structures.
Where this is perceived as a deficiency,
it might be worth considering the choice or design
of a different language that better fits the
particular project context.

Execution of code in parallel may be handled on different
levels of a system. GuStL is designed to implement a
dynamic software system, and thus it provides
parallel execution on process level, where every process
start requires to dynamically determine a free resource,
the loading of code, and setup of control communication.
For static software systems, e.g. embedded control,
things are different in that all parts of the software
needed at runtime are known prior to system start,
i.e. at compile time. Consequently, dynamically loading code
is not necessary and even communication layout between processes
may be arranged at compile time. There are examples that
allow much more efficient instantiation of parallelity
in a static software system (e.g. \lref{2010hh}),
but the question of how to apply these techniques to a
dynamic software system remains open.

Directly addressing memory is not needed even in the context
of an operating system, as the intended target architecture
provides peripheral access through dedicated operations
instead of memory mapped control registers, see \lref{2016os}.

~

\subsection*{Literature}
\addcontentsline{toc}{section}{Literature}

\def\Lit#1#2{\item {\tt [#1]}
#2}
\begin{list}{}{
  \setlength{\labelwidth}{0mm}
  \setlength{\itemsep}{0ex plus0.2ex}
  \setlength{\leftmargin}{6ex}
  \setlength{\itemindent}{-6ex}
  \setlength{\labelsep}{0mm}}

\Lit{1840ob}{
  Otto Böhtlingk (Ed.):
  ``P$\bar{a}$\d{n}ini's Grammatik'',
  Bonn 1840, Delhi 2001,
  ISBN 81-208-1025-2
}

\Lit{1945jn}{
  John von Neumann:
  ``First Draft of a Report on the EDVAC'',
  Moore School of Electrical Engineering, University of Pennsylvania,
  June~30, 1945
}

\Lit{1959jb}{
  J. W. Backus:
  ``The syntax and semantics of the proposed international algebraic language of the Zürich ACM-GAMM conference'',
  Proc. Internat. Conf. Inf. Proc., UNESCO, Paris,
  June 1959
}

\Lit{1963nb}{
  J. W. Backus, F. L. Bauer, J. Green, C. Katz, J. McCarthy, A. J. Perlis,
  H. Rutishauser, K. Samelson, B. Vauquois, J. H. Wegstein, A. van W{\ij}ngaarden,
  M. Woodger, P. Naur (Ed.):
  ``Revised Report on the algorithmic language ALGOL 60'',
  Numerische Mathematik~4, p.~420-453 (1963)
}

\Lit{1975ed}{
  Edsger W. D{\ij}kstra:
  ``Guarded commands, nondeterminacy, and formal derivation of programs'',
  Comm. ACM~18,~8 (Aug.~1975), p.~453-457,
  EWD472
}

\Lit{1975jw}{
  Kathleen Jensen, Niklaus Wirth:
  ``Pascal: user manual and report'',
  Springer-Verlag, 1975
}

\Lit{1977nw}{
  Niklaus Wirth:
  ``Compilerbau'',
  Teubner, Stuttgart,
  ISBN 3-519-02338-5,
  1977
}

\Lit{1978ch}{
  C. A. R. Hoare:
  ``Communicating sequential processes'',
  The Queen's University, Belfast, Northern Ireland
  Commun. ACM~21,~8 (Aug.~1978), p.~666-677
}

\Lit{1978kr}{
  Brian W. Kernighan, Dennis M. Ritchie:
  ``The C Programming Language (1st ed.)'',
  February 1978,
  Englewood Cliffs, NJ: Prentice Hall, ISBN 0-13-110163-3
}

\Lit{1980rw}{
  Martin Richards, Colin Whitby-Strevens:
  ``BCPL -- the language and its compiler'',
  Cambridge, 1980, ISBN 0-521-28681-6
}

\Lit{1982bg}{
  Friedrich L. Bauer, Gerhard Goos:
  ``Informatik -- eine einführende Übersicht'',
  Teil 1,
  1982,
  Springer,
  ISBN 3-540-11722-9
}

\Lit{1997dw}{
  Jochen Demuth, Stephan Weber, Sönke Kannapinn, Mario Kröplin:
  ``Echte Compilergenerierung -- Effiziente Implementierung einer abge\-schlossenen Theorie'',
  Forschungsbericht des Fachbereichs Informatik 1997/6,
  Technische Universität Berlin
}

\Lit{2001dl}{
  Daan Le{\ij}en:
  ``Division and Modulus for Computer Scientists'',
  University of Utrecht,
  Dept. of Computer Science,
  December 3, 2001
}

\Lit{2003fy}{
  F. Yergeau:
  ``UTF-8, a transformation format of ISO 10646'',
  Request For Comments RFC 3629,
  Network Working Group,
  Alis Technologies,
  November 2003
}

\Lit{2007ir}{
  ITU-T Recommendation Z.100:
  ``Specification and Description Language (SDL)'',
  International Telecommunication Union,
  11/2007
}

\Lit{2009uc}{
  The Unicode Consortium:
  ``The Unicode Standard, Version 5.2.0'',
  Mountain View, CA, 2009,
  ISBN 978-1-936213-00-9
}

\Lit{2010hh}{
  James Hanlon, Simon J. Hollis:
  ``Dynamic generation of parallel computations'',
  Department of Computer Science, University of Bristol, UK,
  in proc. UK Electronics Forum 2010, (pp. 7-17), 2010
}

\Lit{2016os}{
  Oskar Schirmer:
  ``NOP -- A Simple Experimental Processor for Parallel Deployment'',
  Göttingen, 2016
}

\end{list}

\end{document}